\documentclass{Interspeech}
\usepackage[T1]{fontenc}     % Proper font encoding
\usepackage{multirow}
\usepackage{footnote}
\usepackage{url}
\usepackage{threeparttable}

\interspeechcameraready

\title{Analysis of ABC Frontend Audio Systems for the NIST-SRE24}

\author[affiliation={1}]{Sara}{Barahona}
\author[affiliation={2}]{Anna}{Silnova}
\author[affiliation={2}]{Ladislav}{Mo\v{s}ner}
\author[affiliation={2}]{Junyi}{Peng}
\author[affiliation={2}]{Old\v{r}ich}{Plchot}
\author[affiliation={2}]{Johan}{Rohdin}
\author[affiliation={2}]{Lin}{Zhang}
\author[affiliation={2}]{Jiangyu}{Han}
\author[affiliation={2}]{Petr}{Palka}
\author[affiliation={2}]{Federico}{Landini}
\author[affiliation={2}]{Luk\'{a}\v{s}}{Burget}
\author[affiliation={3}]{Themos}{Stafylakis}
\author[affiliation={4}]{Sandro}{Cumani}
\author[affiliation={5}]{Dominik}{Bobo\v{s}}
\author[affiliation={5}]{Miroslav}{Hlava\v{c}ek}
\author[affiliation={5}]{Martin}{Kodovsky}
\author[affiliation={5}]{Toma\v{s}}{Pavli\v{c}ek}

\affiliation{${}^{1}$AUDIAS}{Universidad Autónoma de Madrid,Spain \quad ${}^{2}$Brno University of Technology}{Czechia}
\affiliation{}{${}^{3}$Athens University of Economics and Business | Omilia | Archimedes AI/Athena RC}{Greece} 
\affiliation{}{${}^{4}$Politecnico di Torino, Italy \quad ${}^{5}$Phonexia}{Czechia}

\email{sara.barahona@uam.es, \{isilnova, imosner, pengjy, iplchot, rohdin\}@fit.vut.cz}

\usepackage{comment}

\begin{document}

\maketitle

\begin{abstract}
    
    
    %This paper presents a comprehensive analysis of the embedding extractors developed by the XXX team for the NIST Speaker Recognition 2024 Evaluation.  Addressing the evaluation's core audio-track challenges (cross-source and cross-lingual target/non-target trials, enrollment duration variability and shorter test segments), we investigate solutions for both fixed and open conditions. We explored different frontends for the fixed condition including ResNet backbones followed by xi-approach, alongside the state-of-the-art ReDimNet model. In the open condition, given the lack of data constraints and the multilingual nature of the task, we also investigated the multilingual pre-trained model XLS-R and explored the VoxBlink2 dataset, containing over 110k speakers across multiple languages.  Motivated by the strong generalization capabilities of VoxBlink2-trained models, we analyze the impact of increasing training segment lengths and assess resampling effects on the original data domain.

    We present a comprehensive analysis of the embedding extractors (frontends) developed by the ABC team for the audio track of NIST SRE 2024. We follow the two scenarios imposed by NIST: using only a provided set of telephone recordings for training (fixed) or adding publicly available data (open condition). Under these constraints, we develop the best possible speaker embedding extractors for the pre-dominant conversational telephone speech (CTS) domain. We explored architectures based on ResNet with different pooling mechanisms, recently introduced ReDimNet architecture, as well as a system based on the XLS-R model, which represents the family of large pre-trained self-supervised models. In open condition, we train on VoxBlink2 dataset, containing 110 thousand speakers across multiple languages. We observed a good performance and robustness of VoxBlink-trained models, and our experiments show practical recipes for developing state-of-the-art frontends for speaker recognition.
\end{abstract}

\section{Introduction}

The major body of speaker verification (SV) research focuses on 16 kHz datasets with audio excerpts extracted from (e.g., YouTube) video clips. The VoxCeleb datasets~\cite{nagrani17_interspeech,chung2018voxceleb2} represented a great milestone in pushing the verification performance forward by providing training data comprising thousands of speakers. This initiative has been recently followed by other works, releasing large-scale corpora comprising tens of thousands of speakers~\cite{voxblink2} and fostering research in the aforementioned domain.

On the other hand, speaker verification in telephony (characterized by various codecs and 8 kHz sampling rate), which is crucial in many real-world applications, seems to be somewhat overlooked in the community. Fortunately, NIST Speaker Recognition Evaluations have consistently pushed the boundaries of speaker recognition technology for decades, focusing especially on challenges related to detecting speakers in noisy conversational telephone speech (CTS) (but also audio from video, AfV, and multi-modal data). The latest edition (SRE24) was no exception. It included a mandatory audio-only track, as well as optional visual-only and audio-visual tracks that require the integration of speech, image, and video data for speaker identification. The main audio track deals with cross-source (CTS vs. AfV) and cross-lingual target and non-target trials. They were drawn from a new multilingual corpus, TELVID-Tunis. As the name suggests, it focuses on non-mainstream languages: Tunisian Arabic, North African French, and accented English. On top of non-standard languages, this latest edition included some novelties such as enrollment duration variability (10, 30, or 60 seconds), shorter test segments (ranging approximately from 5 seconds to 60 seconds), and multi-speaker enrollment data with diarization annotations. 

The evaluation proposes two training paradigms: fixed and open. The fixed condition constrains participants to use only organizer-provided datasets, including a CTS superset~\cite{SupersetCTS}, NIST SRE 2016~\cite{Sadjadi2017The2N} and 2021 Evaluation~\cite{sadjadi22b_odyssey_nist2021}, and Janus Multimedia set~\cite{sell2018audio}. Such a constraint poses challenges as most of the data comprises CTS only (while the evaluation data contains both CTS and AfV). Moreover, the majority of recordings of the main training data, CTS Superset, are spoken in English (despite containing more than 50 languages), overweighting one of the evaluation languages. In contrast, the open condition allows participants to explore how unlimited additional training data affects system performance.

In this paper, our goal is to build strong frontends for speaker verification following the NIST SRE24 rules. We strive to achieve this by the following means:
\begin{itemize} 
\item \textbf{Variability of frontends}: As per empirical evidence detailed in~\cite{backend_analysis}, diverse frontends tend to be complementary, which provides benefits in fusion. Therefore, we intend to explore ResNet architectures~\cite{he2016resnet} that have proven strong for speaker embedding extraction in numerous evaluations~\cite{huh2024_vox-challenges, sadjadi22_sre21} and challenge the recent model, ReDimNet~\cite{yakovlev24-redimnet}, that was shown to provide state-of-the-art results on the VoxCeleb trial lists and promised strong generalizability.

\item \textbf{Alternative pooling method}: Inspired by consistent improvements in results on the previous NIST evaluation data provided by considering the uncertainty of estimates, we use the xi-vector pooling style within embedding extractors~\cite{lee2021xi}.

\item \textbf{Large-scale out-of-domain dataset}: As noted before, advances in speaker verification on audio data from videos lead to collecting large-scale datasets. We aim to explore and exploit one of them, VoxBlink2~\cite{voxblink2}, in the non-constrained track. Moreover, we will focus on how such out-of-domain datasets can be beneficially used in the context of various applications.

\item \textbf{SSL model}: Foundational models trained in a self-supervised way also attracted attention in speaker verification by providing strong results while requiring a shorter time to fine-tune (compared to training embedding extractors from scratch)~\cite{peng23-mhfa}. Therefore, we aim to explore them in the challenging data.

\end{itemize}

\section{Proposed method}

\subsection{Training data and augmentations}
For the fixed condition, we used the NIST CTS Superset~\cite{SupersetCTS} to train the embedding extractors. To enhance the robustness of the model, we implemented acoustic data enhancement using the Kaldi toolkit \cite{povey2011kaldi}, incorporating noise from the MUSAN database \cite{musan2015} and room impulse responses from the RIR database~\cite{ko2017RIRs}. However, we specifically excluded MUSAN's Babble noise and Music components to comply with fixed track regulations. Prior to model training, non-speech segments were removed through a Kaldi-style energy-based VAD Voice Activity Detection (VAD) system.

The open condition, free from data restrictions, presents an opportunity to leverage large out-of-domain datasets. The largest publicly available speaker verification dataset to date is VoxBlink2 \cite{voxblink2}, which consists of audios from YouTube videos belonging to 111,284 speakers, significantly surpassing the widely used VoxCeleb \cite{chung2018voxceleb2}. However, since this data does not inherently match the CTS domain, we experimented with downsampling the audio to 8kHz while applying GSM codec to 50\% of the data via Sox\footnote{\href{https://sourceforge.net/projects/sox/}{https://sourceforge.net/projects/sox/}} aiming to simulate the telephone channel.

%explored different domain adaptation techniques, including downsampling the audio to 8\,kHz and introducing GSM codec with a 50\% probability using Sox\footnote{\href{https://sourceforge.net/projects/sox/}{https://sourceforge.net/projects/sox/}}, which better approximates real-world telephone conditions.  

\subsection{Frontend systems}
\label{sec:frontend}
In this section, we present the different embedding extractors explored for facing both fixed and open conditions. For reproducibility, we also describe the training setup, implemented using the WeSpeaker toolkit~\cite{wang2023wespeaker,wang2024advancing}. Most of our embedding extractors follow the VoxCeleb recipe, employing all the suggested hyperparameters for the training, which consists of two stages,  both aimed at minimizing the AAM-Softmax loss~\cite{deng2019arcface}. The first stage involves training for 150 epochs with a 2-second segment length and employing a scale of 32 for the AAM loss. Initially, no margin is applied, but between epochs 20 and 40, the margin is gradually increased from 0 to 0.2, and this value is maintained for the remainder of the training. The learning rate scheduler uses a 6-epoch warm-up, linearly increasing the rate from 0 to its highest value (0.1) and then exponential decrease to 5e-5 for the rest of the training. The second stage, so-called large-margin fine-tuning, involves further training for 10 more epochs, employing a larger segment length (10 seconds) and also a larger margin value of 0.5, that remains fixed.

\begin{table*}[th!]
  \caption{Comparison of single frontend systems employing cosine scoring on the SRE 2024 development and evaluation sets for both fixed and open train conditions.}
  \label{tab:result-frontend}
  \centering
  \begin{tabular}{ccccccc}
    \toprule
    & & \multicolumn{2}{c}{SRE24 dev} & \multicolumn{2}{c}{SRE24 eval} \\
     \textbf{Condition} & \textbf{Frontend} & $C_{primary}$ & \textbf{EER (\%)} & $C_{primary}$ & \textbf{EER (\%)} & \textbf{FLOPs (G)} \\
      \midrule
      \multirow{4}{*}{Fixed} & XI-ResNet-34 & 0.688 & 13.91  & 0.747 & 14.40 & 551 \\
       & XI-ResNet-152 & 0.615 & \textbf{10.16} & 0.695 & 10.41 & 176 \\
      & XI-ResNet-221 & \textbf{0.597} & 10.26 & \textbf{0.683} & \textbf{10.18} & 254 \\
      & ReDimNet-B3 & 0.728  & 14.70 & 0.784 & 14.33 & \textbf{71} \\
      \midrule
      \multirow{2}{*}{Open} & ResNet-152-VB & \textbf{0.522} & \textbf{9.31} & \textbf{0.562} & \textbf{7.59} & 219 \\
      & XLS-R & 0.666 & 12.02 & 0.681 & 11.69 & 270 \\
      \bottomrule
  \end{tabular}
\end{table*}

% The training followed the VoxCeleb recipe of WeSpeaker for ResNets utilizing all of the suggested hyperparameters. Training minimizes the AAM objective~\cite{deng2019arcface} with the margin set to 0.2 and the scale set to 32. For the first 20 epochs of training, the margin is set to 0 and then gradually increased from 0
% to 0.2 over a course of 20 epochs. Finally, it is fixed for the rest of the training until epoch 150 is reached. The learning rate scheduler is set as in the original recipe - warming up the learning rate from 0 to its highest value (0.1) for 6 epochs and then exponentially decreasing it
% to 5e-5 for the rest of the training. The training was performed using segments of 2 seconds duration. After 150 epochs, we increased the length of the training segments to 10 seconds and continued training for 10 more epochs with the fixed margin and learning rate.

\textbf{\textit{XI-ResNet}}: For the fixed condition, we explored a set of ResNet models~\cite{he2016resnet}, which have shown a strong performance in speaker recognition under challenging conditions. Specifically, we explored ResNet34, ResNet152 and ResNet221 architectures. Our key innovation lies in replacing the standard temporal statistic pooling (TSTP) layer with the xi-vector approach \cite{lee2021xi}. This method integrates uncertainty estimation by incorporating the Bayesian formulation of the linear Gaussian model (i-vector) directly into the pooling layer of the speaker-embedding neural network. 

For the XI-ResNet152, we experimented with different modifications on the aforementioned setup including longer training segments of 3 seconds, speed perturbation was turned off for this experiment, and instead of running the training for 150 epochs followed by 10 additional epochs with 10s training examples, the first stage this time lasted 130 epochs and the second one 5 epochs.

% \textbf{\textit{XI-ResNet-221}}: {\color{red}[[Should we include it? It is rather controversial]]} We also explore using an even larger encoder for the ResNet models, following the same training procedure as in the above system. However, for this system, we skipped the second stage of fine-tuning on longer segments. \textbf{perform second stage?}.

\textbf{\textit{ReDimNet-B3}}: As an alternative to ResNet models, we explore the recently-proposed Reshape Dimensions Network (ReDimNet) \cite{yakovlev24-redimnet} architecture under fixed conditions. Although ReDimNet has achieved state-of-the-art results in the VoxCeleb benchmark, it has not yet been applied to the NIST domain. ReDimNet integrates 1D and 2D convolutional blocks by reshaping dimensionality between feature map representations in a single model. We hypothesize that by combining these two blocks, it could better capture the complex temporal and spectral variations present in telephone speech, including those introduced by limited bandwidth and channel noise, compared to ResNets, which predominantly use 2D convolutions. Specifically, we selected the B3 version based on our empirical experience, as it will be shown in Section 4. During the large-margin fine-tuning, we trained for 5 epochs employing six-second segments.
% The final model was trained in two stages, with short and long segments. The hyperparameters for the first follow those described for XI-ResNet-34. In the second stage (so-called large-margin finetuning), the model was further trained for five epochs on six-second segments, optimizing AAM with a margin increased to 0.5. The learning rate decreased exponentially from 1e-4 to 2.5e-5.

% \section{Open-condition}
% \subsection{Training data}
% In addition to the data employed for the fixed condition, we experiment with the following external data, which we downsampled to 8kHz:
% \begin{itemize}
%     \item VoxBlink2 dataset
%     \item VoxCeleb2 development set \textbf{backend?}
%     \item NIST SRE 20218 development and evaluation sets \textbf{backend?}
% \end{itemize}

% \subsection{Frontend systems}
% \label{section:open-frontends}
\textit{\textbf{ResNet-152-VB}}: To leverage the unconstrained data paradigm of the open condition, we explored the VoxBlink2 dataset by training a ResNet152 model. While the xi-approach demonstrated notable efficacy in the fixed condition, we employed the conventional temporal statistic pooling for this model to systematically isolate and evaluate the impact of incorporating the VoxBlink2 data corpus. For feature extraction, we computed 80-dimensional log Mel-filterbank energy features. Following the initial training on VoxBlink2, we performed large-margin fine-tuning on the CTS Superset. As detailed in Section 4, during this stage, we explored varying segment durations to assess their influence on performance across the enrollment and test conditions introduced in SRE24.
% It was trained on 2-second segments for 150 epochs using AAM-Softmax loss with a scale factor of 32. The margin parameter is gradually increased from 0 to 0.2 between epochs 20 and 40. We applied an exponential decay to the learning rate after a 6-epoch warm-up, with a maximum rate of 0.1 and a final rate of 5e-5.
% Following training on VoxBlink2, we performed Large-Margin Fine-Tuning on the CTS Superset. Specifically, the model was further trained for 10 epochs employing lower learning rates than in the previous phase, starting from 1e-4 and exponentially decreasing to 2.5e-5.

\textbf{\textit{XLS-R}}: In the open condition, we also made use of a foundation model pre-trained in a self-supervised way, trying to confirm/contradict the benefits shown in the context of the VoxCeleb data \cite{peng23-mhfa}. Specifically, we opted for XLS-R \cite{babu22-xlsr} as it was pre-trained on 436K hours of multilingual data comprising (at least dialects of) languages in the evaluation set. A notable advantage of this model is that a subset of pre-training examples is sampled at 8 kHz and contains telephone speech. In the fine-tuning stage, we appended a multi-head factorized attention (MHFA) backend~\cite{peng23-mhfa} to the pre-trained XLS-R 300M and fine-tuned both components on upsampled CTS Superset recordings, optimizing an AAM Softmax loss (with a scale of 32 and a margin of 0.2). MHFA is a lightweight attention-based embedding extractor compatible with various backbones comprising transformer encoder blocks since it employs per-frame representations at various levels of models (arguably rich in different information, e.g., phonetic or speaker-related). MHFA comprised 64 heads and produced 256-dimensional embeddings. The learning rate decreased exponentially from 1e-2 to 4.4e-3 over the course of 30 epochs. The pre-trained weights of XLS-R were updated using a learning rate scaled down by a factor of 0.08 compared to MHFA.

\section{Experimental Results}
Focusing on comparing and analyzing the effects of different embedding extractors, we employed cosine scoring as our classifier, a natural choice given our optimization of the AAM loss. To isolate the effects of different frontend systems, we applied a consistent preprocessing pipeline consisting of centering, dimensionality reduction using Linear Discriminant Analysis (LDA) and length normalization of embeddings. Specifically, we intentionally omit additional pre-processing or calibration techniques to focus on the embedding extractor itself.  However, for models trained on the VoxBlink2 dataset, we found it beneficial to exclude the LDA step. Performance is evaluated using Equal Error Rate (EER) and $C_{primary}$, as defined by the SRE24 evaluation plan \cite{NIST24}. Computational complexity is assessed using Floating Point Operations per second (FLOPs).

% This involved two applications of Nuisance Attribute Projection (NAP). The first NAP step removed speaker gender information, while the second addressed source variability (cts vs. afv). We estimated the mean for the former from the CTS Superset, while the latter was estimated on afv part of NIST SRE 2021. Finally, the embeddings were centered, LDA dimensionality reduction, and length normalization.  

\subsection{Fixed Systems}
In Table~\ref{tab:result-frontend}, we show the results for the different frontends explored for the fixed condition. While ReDimNet has achieved state-of-the-art results on the VoxCeleb benchmark, ResNet-based backbones consistently outperformed it across both the development and evaluation sets of the SRE24 challenge. This performance advantage of ResNet was observed despite ReDimNet's lower computational complexity, achieving the lowest FLOP value. As expected, among the ResNet models, the larger XI-ResNet-221 achieves the best performance. However, its gains over the mid-sized XI-ResNet-152 are not drastic, indicating that the benefits of scaling up the model size may be limited beyond a certain point.

The development of the ReDimNet frontend involved exploring different model sizes, as detailed in Table \ref{tab:results-redimnet}. Specifically, we evaluated configurations B0, B2, B3, and B6, on the development set to identify the optimal model size. The B3 configuration yielded the best performance, suggesting that further increases in model complexity yielded minimal gains in our domain. Given that even this optimized ReDimNet configuration did not surpass the performance of even the smallest ResNet model, only the B3 configuration was included in our final submission.

\subsection{Open Systems}
Results for the open condition system are also shown in Table~\ref{tab:result-frontend}. The use of VoxBlink2 dataset during the first stage of the training has an enormous positive impact on the results, considerably improving performance metrics. This improvement is primarily attributed to the dataset's extensive speaker diversity, which strengthens the model's generalization capabilities and results in a notably lower EER on the evaluation set. 

In contrast, our attempt to develop a robust cross-lingual system using the pre-trained multi-lingual XLS-R model did not yield the expected improvements, obtaining a performance degradation compared to the fixed condition systems. Given this model was trained with a set of 8\,kHz data and contained telephone speech, results suggest that further fine-tuning strategies should be studied.

\begin{table}[t!]
  \caption{Ablation study of different ReDimNet configurations over the SRE development set.}
  \label{tab:results-redimnet}
  \centering
  \begin{tabular}{ccc}
    \toprule
     \textbf{Frontend} & $C_{primary}$ & \textbf{EER (\%)} \\
      \midrule
      ReDimNet-B0 & 0.943 & 27.47 \\
      ReDimNet-B2 & 0.783 & 15.48 \\
      ReDimNet-B3 & \textbf{0.728} & \textbf{14.70}  \\
      ReDimNet-B6 & 0.777 & 18.76 \\
    \bottomrule
  \end{tabular}
\end{table}

\subsection{Effects of resampling Voxblink2 dataset}
\begin{table*}[th]
\caption{Effect of Voxblink2 sampling rate on SRE24 and VoxCeleb1 datasets. VoxCeleb1 results are only shown for models trained during the first stage, employing no margin-finetuning on a different dataset. The impact of fine-tuning ResNet152-VB on different segment lengths is also shown.}
\label{tab:voxblink-models}
\centering
\begin{threeparttable}
\begin{tabular}{crcccccccc}
\toprule
& & & \multicolumn{2}{c}{SRE24 dev} & \multicolumn{2}{c}{SRE24 eval} & \multicolumn{3}{c}{VoxCeleb1} \\
\textbf{Data} & \textbf{Frontend} & \textbf{FT length} & $C_{prim.}$ & \textbf{EER (\%)} & $C_{prim.}$ & \textbf{EER (\%)} & \textbf{O} & \textbf{E} & \textbf{H} \\
\midrule
16 kHz & SimAM-ResNet34-ASP\tnote{2} & 10s & 0.536 & 9.11 & 0.611 & 8.10 & 1.11 & 1.17 & 2.24\\
16 kHz & SimAM-ResNet100-ASP\tnote{2} & 10s & 0.590 & 9.75 & 0.634 & 8.06 & \textbf{0.76} & \textbf{0.89} & \textbf{1.76}\\
16 kHz + 8 KHz $\uparrow$ & ResNet152-VB & 10s & 0.541 & \textbf{8.99} & 0.582 & \textbf{7.53} & 1.65 & 1.37 & 2.73 \\
8 kHz & ResNet152-VB & 10s & \textbf{0.522} & 9.31 & \textbf{0.562} & 7.59 & 2.42 & 2.15 & 4.32 \\
\midrule
\multirow{3}{*}{8kHz} & \multirow{3}{*}{ResNet152-VB} & 6s & 0.544 & 10.06 & 0.641 & 8.51 & - & - & -\\
& & 20s & 0.519 & 8.81 & 0.534 & 7.25 & - & - & -\\
& & 40s & \textbf{0.482} & \textbf{7.93} & \textbf{0.491} & \textbf{6.47} & - & - & - \\
\bottomrule
\end{tabular}
\begin{tablenotes}
    \scriptsize 
    \item[2] \url{https://github.com/wenet-e2e/wespeaker/blob/master/docs/pretrained.md}.
\end{tablenotes}
\end{threeparttable}
\end{table*}

When training our ResNet152 over the VoxBlink2 dataset, we initially downsampled the audio to 8 kHz to align with the characteristics of our fine-tuning domain, the CTS Superset. Pre-training on this substantial corpus of domain-adapted data yielded our best-performing system, demonstrating strong generalization on the SRE24 evaluation set. However, this 8 kHz pre-trained model exhibited limited generalization to other domains, as evidenced by the VoxCeleb1 results presented in Table~\ref{tab:voxblink-models}.

To further analyze the impact of resampling data in the different domains, we explored training with a combined dataset comprising both the original 16 kHz audio and the resampled 8 kHz data. The previously downsampled 8 kHz audio, which also had the GSM codec applied randomly to 50\% of segments, was upsampled back to 16 kHz. This approach exposed the model to both original and simulated telephone speech. We trained ResNet152 with exactly the same parameters as detailed in Section~\ref{sec:frontend}. However, due to the doubled dataset size and that each epoch iterates over the whole dataset, we trained for 80 epochs to approximate the total number of training iterations used previously. As shown in Table~\ref{tab:voxblink-models}, this combined-data model achieved comparable performance to the 8 kHz version on the SRE24 setup, even outperforming it in terms of EER. Additionally, incorporating the original 16 kHz data significantly improved generalization to other real domains, resulting in a performance gain on the VoxCeleb dataset.

For a comprehensive comparison, we also included available pre-trained VoxBlink2 models from WeSpeaker in Table~\ref{tab:voxblink-models}. While these models are also ResNet-based, they are not directly comparable due to architectural differences: they incorporate Simple Attention Modules (SimAM)~\cite{simAM-yang21o} within the ResNet blocks and utilize Attentive Statistics Pooling (AST)~\cite{okabe18_interspeech} as aggregation function.
While these 16kHz systems demonstrated strong performance, they did not outperform our ResNet152-VB models on the SRE24 evaluation set.
% Although the SimAM-ResNet100-ASP model achieves the best performance on the SRE24 development set, our models trained with  outperform them on the evaluation set. 
% %This suggests our data augmentation strategy, simulating realistic telephone effects, is more effective for generalization than architectural refinements, highlighting the importance of training data characteristics for robust speaker recognition.  
Nevertheless, using state-of-the-art pre-trained embedding extractors such as SimAM-ResNet100-ASP and its fine-tuned version on the CTS Superset is a very good alternative to our training approaches with 8kHz or 16kHz hybrid data. These systems are very compelling when we aim for a universal system capable of operating both in the telephone and wideband domains.

% \begin{figure}[t]
%   \centering
%   \includegraphics[width=\linewidth]{voxceleb1-voxvlink.jpg}
%   \caption{Results with ResNet-based models trained on the VoxBlink2 dataset over the VoxCeleb1 trials. Results are shown for models trained during the first stage, employing no margin-finetuning on a different dataset.}
%   \label{fig:speech_production}
% \end{figure}

\begin{figure}[t]
  \centering
  \includegraphics[width=\linewidth,trim={4cm 8.5cm 4.5cm 9cm},clip]{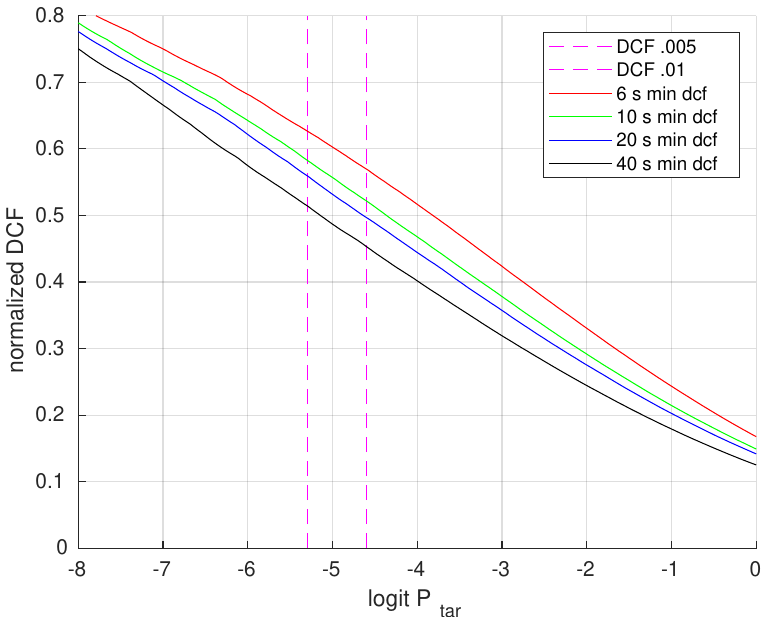}
  \caption{DCF plots for ResNet152-VB fine-tuned on segments of different lengths.}
  \label{fig:len_dcf_plots}
\end{figure}

% \begin{figure}[t]
%   \centering
%   \includegraphics[width=0.9\linewidth,trim={3.5cm 6.5cm 3.5cm 7cm},clip]{img/len_comp_det.pdf}
%   \caption{{\color{red}Not sure if we will use it but I am adding it now for you to see the comparison of the systems fine-tuned with different lengths of segments}.}
%   \label{fig:len_det_plots}
% \end{figure}

\subsection{Impact of segment length in the fine-tuning process}
Given that one of the key challenges introduced in SRE24 was the presence of shorter test segments and variability in enrollment durations, we investigated the impact of different segment lengths during the fine-tuning stage on the CTS Superset using our ResNet152-VB model. We systematically increased the segment length up to 40 seconds, observing a consistent improvement in performance. While both the development and evaluation sets exhibited gains, the evaluation set benefited the most, achieving a 23.98\% reduction in EER. In terms of $C_{primary}$, the extension of the segment length also contributed to a better generalization, closing the gap between the evaluation and development results. 
%Specifically, we observed a 20.39\% reduction in minDCF when using $P_{tar}$ = 0.01 and a 17.91\% reduction for $P_{tar}$ = 0.005.

The aforementioned $C_{primary}$ results depend on the operating points chosen by the organizers. They correspond to a specific application of the verification systems. In order to provide insight into the expected performance in various applications, we show DCF plots in Figure \ref{fig:len_dcf_plots}. Not only do we observe improvement stemming from longer training segments for both operating points of interest (marked by vertical dashed lines), but it is consistent across a wide range of operating points. We note that we did not partition scores in any way when computing minDCF metric here.

% \begin{figure}[t]
%   \centering
%   \includegraphics[width=\linewidth]{vo.eps}
%   \caption{Effect of increasing the segment length during the large-margin fine-tuning stage on the CTS dataset for the ResNet152-VB model.}
%   \label{fig:speech_production}
% \end{figure}

% % 
% \begin{align}
%   x(t) &= s(t') \nonumber \\ 
%        &= s(f_\omega(t))
% \end{align}
% % 
% where \(f_\omega(t)\) is a special warping function. Equation \ref{equation:eq2} is a little more complicated.
% % 
% \begin{align}
%   f_\omega(t) &= \frac{1}{2 \pi j} \oint_C 
%   \frac{\nu^{-1k} \mathrm{d} \nu}
%   {(1-\beta\nu^{-1})(\nu^{-1}-\beta)}
%   \label{equation:eq2}
% \end{align}
% % 

% \begin{figure}[t]
%   \centering
%   \includegraphics[width=\linewidth]{figure.pdf}
%   \caption{Schematic diagram of speech production.}
%   \label{fig:speech_production}
% \end{figure}

\section{Conclusion}

This paper targeted speaker verification in a challenging mixture of telephony and audio-from-video speech. With the aim of building a strong frontend, we explored various architectures, pooling methods, and models pre-trained in a self-supervised way.

What we consider the most prominent contribution is the analysis of using large-scale (not necessarily in-domain) datasets to obtain noteworthy performance and generalizability. First, we showed substantial improvements from models pre-trained on (potentially downsampled and GSM-augmented) VoxBlink2 and fine-tuned on CTS Superset. Bearing in mind the length of enrollment and test recordings, we gradually increased the duration of fine-tuning segments while consistently improving performance across a wide range of operating points. Finally, our ResNet152-VB model targeted the domain of the NIST SRE24 evaluation data. However, it lacks strong generalization ability, which was tested on the VoxCeleb data. We showed that a comparable performance on the evaluation data and considerable generalization ability can be achieved when pre-training the model on a mixture of original 16 kHz data and its copy, which was subject to downsampling with the optional application of GSM codec and followed by upsampling back to 16kHz.  

We plan to release our models pre-trained on VoxBlink2 on WeSpeaker website to complement already available models. Using such pre-trained models can save the research community valuable computing resources.
% \begin{figure}[t!]
%   \centering
%   \includegraphics[width=\linewidth]{ablation_study_lengthVB_minDCF.jpg}
%   \caption{Impact of increasing segment length during large-margin fine-tuning stage on the CTS dataset for the ResNet152-VB model, with minDCF evaluated at $P_{tar}$ = 0.001 (colored bars) and  $P_{tar}$ = 0.005 (transparent bars).}
%   \label{fig:vb2-length}
% \end{figure}

\section{Acknowledgements}
This work was partly supported by project PID2021-125943OB-I00, MCIN/AEI/10.13039/501100011033/FEDER, UE from the
Spanish Ministerio de Ciencia e Innovacion,
Fondo Europeo de Desarrollo Regional

% \ifinterspeechfinal
%      The Interspeech 2025 organisers
% \else
%      The authors
% \fi
% would like to thank ISCA and the organising committees of past Interspeech conferences for their help and for kindly providing the previous version of this template.

\bibliographystyle{IEEEtran}
\bibliography{mybib}

\end{document}